\documentclass[12pt,a4paper]{article}

\usepackage{ifthen} 
\newboolean{pdflatex}
\setboolean{pdflatex}{true} 

\newboolean{articletitles}
\setboolean{articletitles}{true} 

\newboolean{uprightparticles}
\setboolean{uprightparticles}{false} 

\def\paperauthors{LHCb collaboration} 
\def\paperasciititle{Search for the decay B0 -> phi phi} 
\def\papertitle{Search for the decay $B^0 \rightarrow \phi \phi$}
\def\paperkeywords{{High Energy Physics}, {LHCb}} 
\def\papercopyright{\the\year\ CERN for the benefit of the LHCb collaboration} 
\def\paperlicence{CC BY 4.0 licence}
\def\paperlicenceurl{https://creativecommons.org/licenses/by/4.0/}

\newif\ifEnableSectionTOCLinks
\EnableSectionTOCLinksfalse

\usepackage[top=1in, bottom=1.25in, left=1in, right=1in]{geometry}

\usepackage{microtype}
\usepackage{lineno}  
\usepackage{xspace} 
\usepackage{caption}

\usepackage{graphicx}  
\usepackage{color}
\usepackage{colortbl}
\graphicspath{{./figs/}} 
\usepackage{amsmath} 
\usepackage{amssymb}
\usepackage{amsfonts}
\usepackage{upgreek} 
\newcommand*\patchAmsMathEnvironmentForLineno[1]{%
\expandafter\let\csname old#1\expandafter\endcsname\csname #1\endcsname
\expandafter\let\csname oldend#1\expandafter\endcsname\csname
end#1\endcsname
 \renewenvironment{#1}%
   {\linenomath\csname old#1\endcsname}%
   {\csname oldend#1\endcsname\endlinenomath}%
}
\newcommand*\patchBothAmsMathEnvironmentsForLineno[1]{%
  \patchAmsMathEnvironmentForLineno{#1}%
  \patchAmsMathEnvironmentForLineno{#1*}%
}
\AtBeginDocument{%
\patchBothAmsMathEnvironmentsForLineno{equation}%
\patchBothAmsMathEnvironmentsForLineno{align}%
\patchBothAmsMathEnvironmentsForLineno{flalign}%
\patchBothAmsMathEnvironmentsForLineno{alignat}%
\patchBothAmsMathEnvironmentsForLineno{gather}%
\patchBothAmsMathEnvironmentsForLineno{multline}%
\patchBothAmsMathEnvironmentsForLineno{eqnarray}%
}

\usepackage{hyperxmp}

\usepackage[pdftex,
            pdfauthor={\paperauthors},
            pdftitle={\paperasciititle},
            pdfkeywords={\paperkeywords},
            pdfcopyright={Copyright (C) \papercopyright},
            pdflicenseurl={\paperlicenceurl}]{hyperref}

\usepackage[colorinlistoftodos,textsize=scriptsize]{todonotes}

\usepackage[bottom,flushmargin,hang,multiple]{footmisc}

\usepackage[all]{hypcap}

\def\lhcb   {\mbox{LHCb}\xspace}

\def\babar  {\mbox{BaBar}\xspace}

\def\MagUp {\mbox{\em Mag\kern -0.05em Up}\xspace}

\ifthenelse{\boolean{uprightparticles}}%
{

 \def\Pmu         {\ensuremath{\upmu}\xspace}

 \def\Ppi         {\ensuremath{\uppi}\xspace}

 \def\Ppsi        {\ensuremath{\uppsi}\xspace}

 \def\PDelta      {\ensuremath{\Delta}\xspace}                 
 \def\PXi         {\ensuremath{\Xi}\xspace}                 
 \def\PLambda     {\ensuremath{\Lambda}\xspace}                 
 \def\PSigma      {\ensuremath{\Sigma}\xspace}                 
 \def\POmega      {\ensuremath{\Omega}\xspace}                 
 \def\PUpsilon    {\ensuremath{\Upsilon}\xspace}
 \let\oldPi\Pi
 \def\PPi         {\ensuremath{\oldPi}\xspace}

 \def\PB      {\ensuremath{\mathrm{B}}\xspace}                 
 \def\PD      {\ensuremath{\mathrm{D}}\xspace}                 
 \def\PJ      {\ensuremath{\mathrm{J}}\xspace}                 
 \def\PK      {\ensuremath{\mathrm{K}}\xspace}                 
 \def\PZ      {\ensuremath{\mathrm{Z}}\xspace}                 
 \def\Pb      {\ensuremath{\mathrm{b}}\xspace}                 
 \def\Pc      {\ensuremath{\mathrm{c}}\xspace}

 \def\Pp      {\ensuremath{\mathrm{p}}\xspace}                 

 \def\Ps      {\ensuremath{\mathrm{s}}\xspace}

 \def\thebaroffset{0.0em}
}
{

 \def\Pmu         {\ensuremath{\mu}\xspace}

 \def\Ppi         {\ensuremath{\pi}\xspace}

 \def\Ppsi        {\ensuremath{\psi}\xspace}                 
                  
 \mathchardef\PDelta="7101
 \mathchardef\PXi="7104
 \mathchardef\PLambda="7103
 \mathchardef\PSigma="7106
 \mathchardef\POmega="710A
 \mathchardef\PUpsilon="7107
 \mathchardef\PPi="7105
 \def\PB      {\ensuremath{B}\xspace}                 
 \def\PD      {\ensuremath{D}\xspace}                 
 \def\PJ      {\ensuremath{J}\xspace}                 
 \def\PK      {\ensuremath{K}\xspace}                 
 \def\PZ      {\ensuremath{Z}\xspace}                 
 \def\Pb      {\ensuremath{b}\xspace}                 
 \def\Pc      {\ensuremath{c}\xspace}

 \def\Pp      {\ensuremath{p}\xspace}                 

 \def\Ps      {\ensuremath{s}\xspace}

 \def\thebaroffset{0.18em}
}
\newcommand{\offsetoverline}[2][\thebaroffset]{\kern #1\overline{\kern -#1 #2}}%

\makeatletter
\ifcase \@ptsize \relax% 10pt
  \newcommand{\miniscule}{\@setfontsize\miniscule{4}{5}}% \tiny: 5/6
\or% 11pt
  \newcommand{\miniscule}{\@setfontsize\miniscule{5}{6}}% \tiny: 6/7
\or% 12pt
  \newcommand{\miniscule}{\@setfontsize\miniscule{5}{6}}% \tiny: 6/7
\fi
\makeatother

\DeclareRobustCommand{\optbar}[1]{\shortstack{{\miniscule (\rule[.5ex]{1.25em}{.18mm})}
  \\ [-.7ex] $#1$}}

\def\mumu       {{\ensuremath{\Pmu^+\Pmu^-}}\xspace}

\def\squark    {{\ensuremath{\Ps}}\xspace}

\def\cquark    {{\ensuremath{\Pc}}\xspace}

\def\bquark    {{\ensuremath{\Pb}}\xspace}

\def\pion   {{\ensuremath{\Ppi}}\xspace}

\def\pim    {{\ensuremath{\pion^-}}\xspace}

\def\kaon    {{\ensuremath{\PK}}\xspace}
\def\KorKbar {\kern \thebaroffset\optbar{\kern -\thebaroffset \PK}{}\xspace}

\def\Kp      {{\ensuremath{\kaon^+}}\xspace}
\def\Km      {{\ensuremath{\kaon^-}}\xspace}
\def\Kpm     {{\ensuremath{\kaon^\pm}}\xspace}

\def\Kstarz  {{\ensuremath{\kaon^{*0}}}\xspace}

\def\Kstar   {{\ensuremath{\kaon^*}}\xspace}

\def\D       {{\ensuremath{\PD}}\xspace}

\def\DorDbar {\kern \thebaroffset\optbar{\kern -\thebaroffset \PD}\xspace}

\def\Dp      {{\ensuremath{\D^+}}\xspace}
\def\Dm      {{\ensuremath{\D^-}}\xspace}

\def\DpDm    {\ensuremath{\Dp {\kern -0.16em \Dm}}\xspace}

\def\Ds      {{\ensuremath{\D^+_\squark}}\xspace}

\def\B       {{\ensuremath{\PB}}\xspace}

\def\BorBbar {\kern \thebaroffset\optbar{\kern -\thebaroffset \PB}\xspace}
\def\Bz      {{\ensuremath{\B^0}}\xspace}

\def\Bd      {{\ensuremath{\B^0}}\xspace}

\def\BdorBdbar {\kern \thebaroffset\optbar{\kern -\thebaroffset \Bd}\xspace}
\def\Bu      {{\ensuremath{\B^+}}\xspace}

\def\Bp      {{\ensuremath{\Bu}}\xspace}

\def\Bpm     {{\ensuremath{\B^\pm}}\xspace}

\def\Bs      {{\ensuremath{\B^0_\squark}}\xspace}

\def\BsorBsbar {\kern \thebaroffset\optbar{\kern -\thebaroffset \Bs}\xspace}

\def\Bds     {{\ensuremath{\B_{(\squark)}^0}}\xspace}

\def\BdorBs  {\Bds}

\def\jpsi     {{\ensuremath{{\PJ\mskip -3mu/\mskip -2mu\Ppsi}}}\xspace}

\def\Y#1S{\ensuremath{\PUpsilon{(#1S)}}\xspace}

\def\proton      {{\ensuremath{\Pp}}\xspace}

\def\Lz          {{\ensuremath{\PLambda}}\xspace}

\def\LorLbar     {\kern \thebaroffset\optbar{\kern -\thebaroffset \PLambda}\xspace}

\def\Xires       {{\ensuremath{\PXi}}\xspace}

\def\Lb           {{\ensuremath{\Lz^0_\bquark}}\xspace}

\def\Xibz         {{\ensuremath{\Xires^0_\bquark}}\xspace}

\newcommand{\decay}[2]{\mbox{\ensuremath{#1\!\to #2}}\xspace} 

\def\to                 {\ensuremath{\rightarrow}\xspace}

\def\CP                {{\ensuremath{C\!P}}\xspace}

\def\AT#1     {\ensuremath{A_{\mathrm{T}}^{#1}}\xspace}           % 2

\def\C#1      {\ensuremath{\mathcal{C}_{#1}}\xspace}                       % 9
\def\Cp#1     {\ensuremath{\mathcal{C}_{#1}^{'}}\xspace}                    % 7
\def\Ceff#1   {\ensuremath{\mathcal{C}_{#1}^{\mathrm{(eff)}}}\xspace}        % 9  
\def\Cpeff#1  {\ensuremath{\mathcal{C}_{#1}^{'\mathrm{(eff)}}}\xspace}       % 7
\def\Ope#1    {\ensuremath{\mathcal{O}_{#1}}\xspace}                       % 2
\def\Opep#1   {\ensuremath{\mathcal{O}_{#1}^{'}}\xspace}                    % 7

\newcommand{\aunit}[1]{\ensuremath{\text{\,#1}}}       
\newcommand{\tev}{\aunit{Te\kern -0.1em V}\xspace}
\newcommand{\gev}{\aunit{Ge\kern -0.1em V}\xspace}
\newcommand{\mev}{\aunit{Me\kern -0.1em V}\xspace}
\newcommand{\kev}{\aunit{ke\kern -0.1em V}\xspace}
\newcommand{\ev}{\aunit{e\kern -0.1em V}\xspace}
 
\newcommand{\mevc}{\ensuremath{\aunit{Me\kern -0.1em V\!/}c}\xspace}
\newcommand{\gevc}{\ensuremath{\aunit{Ge\kern -0.1em V\!/}c}\xspace}
\newcommand{\mevcc}{\ensuremath{\aunit{Me\kern -0.1em V\!/}c^2}\xspace}
\newcommand{\gevcc}{\ensuremath{\aunit{Ge\kern -0.1em V\!/}c^2}\xspace}
\def\fb   {\ensuremath{\aunit{fb}}\xspace}
\def\invfb   {\ensuremath{\fb^{-1}}\xspace}

\def\ps   {\ensuremath{\aunit{ps}}\xspace}

\newcommand{\stat}{\aunit{(stat)}\xspace}
\newcommand{\syst}{\aunit{(syst)}\xspace}

\newcommand{\chisqip}{\ensuremath{\chi^2_{\text{IP}}}\xspace}

\def\gsim{{~\raise.15em\hbox{$>$}\kern-.85em
          \lower.35em\hbox{$\sim$}~}\xspace}
\def\lsim{{~\raise.15em\hbox{$<$}\kern-.85em
          \lower.35em\hbox{$\sim$}~}\xspace}

\def\sPlot{\mbox{\em sPlot}\xspace}

\def\pt         {\ensuremath{p_{\mathrm{T}}}\xspace}

\def\ptot       {\ensuremath{p}\xspace}

\def\evtgen     {\mbox{\textsc{EvtGen}}\xspace}

\def\geant      {\mbox{\textsc{Geant4}}\xspace}

\def\photos     {\mbox{\textsc{Photos}}\xspace}

\def\pythia     {\mbox{\textsc{Pythia}}\xspace}

\def\tell1  {TELL1\xspace}
\def\ukl1   {UKL1\xspace}

\newcommand{\lhcborcid}[1]{\href{https://orcid.org/#1}{\hspace*{0.1em}\raisebox{-0.45ex}{\includegraphics[width=1em]{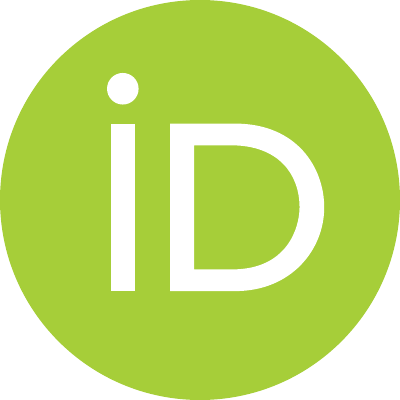}}}}

\hypersetup{
  colorlinks   = true, %Colours links instead of ugly boxes
  urlcolor     = blue, %Colour for external hyperlinks
  linkcolor    = blue, %Colour of internal links
  citecolor    = red   %Colour of citations
}

\ifEnableSectionTOCLinks
    \usepackage[explicit]{titlesec} % to change headings
    
    \let\oldcontentsline\contentsline
    \renewcommand

    \titleformat{\section}{\normalfont\Large\bf}{\hyperlink{tocsection.\thesection}{{\thesection} \parbox[t]{\dimexpr\textwidth-1pc}{#1}}}{1pc}{}

    \titleformat{\subsection}{\normalfont\bf}{\hyperlink{tocsubsection.\thesubsection}{{\thesubsection} \parbox[t]{\dimexpr\textwidth-1pc}{#1}}}{1pc}{}

    \titleformat{name=\section,numberless}[display]{}{}{0pt}{\normalfont\Huge\bfseries #1}
\fi

\usepackage{cite} % Allows for ranges in citations
\usepackage{mciteplus}

\usepackage{longtable} % only for template; not usually to be used in PAPERs

\begin{document}

%%%%%%%%%%%%%%%%%%%%%%%%%
%%%%% Title     %%%%%%%%%
%%%%%%%%%%%%%%%%%%%%%%%%%
\renewcommand{\thefootnote}{\fnsymbol{footnote}}
\setcounter{footnote}{1}

%%%%%%%%%%%%%%%%%%%%%%%%%
%%%%%  TITLE PAGE  %%%%%%
%%%%%%%%%%%%%%%%%%%%%%%%%
\begin{titlepage}
\pagenumbering{roman}

% Header ---------------------------------------------------
\vspace*{-1.5cm}
\centerline{\large EUROPEAN ORGANIZATION FOR NUCLEAR RESEARCH (CERN)}
\vspace*{1.5cm}
\noindent
\begin{tabular*}{\linewidth}{lc@{\extracolsep{\fill}}r@{\extracolsep{0pt}}}
\vspace*{-1.5cm}\mbox{\!\!\!\includegraphics[width=.14\textwidth]{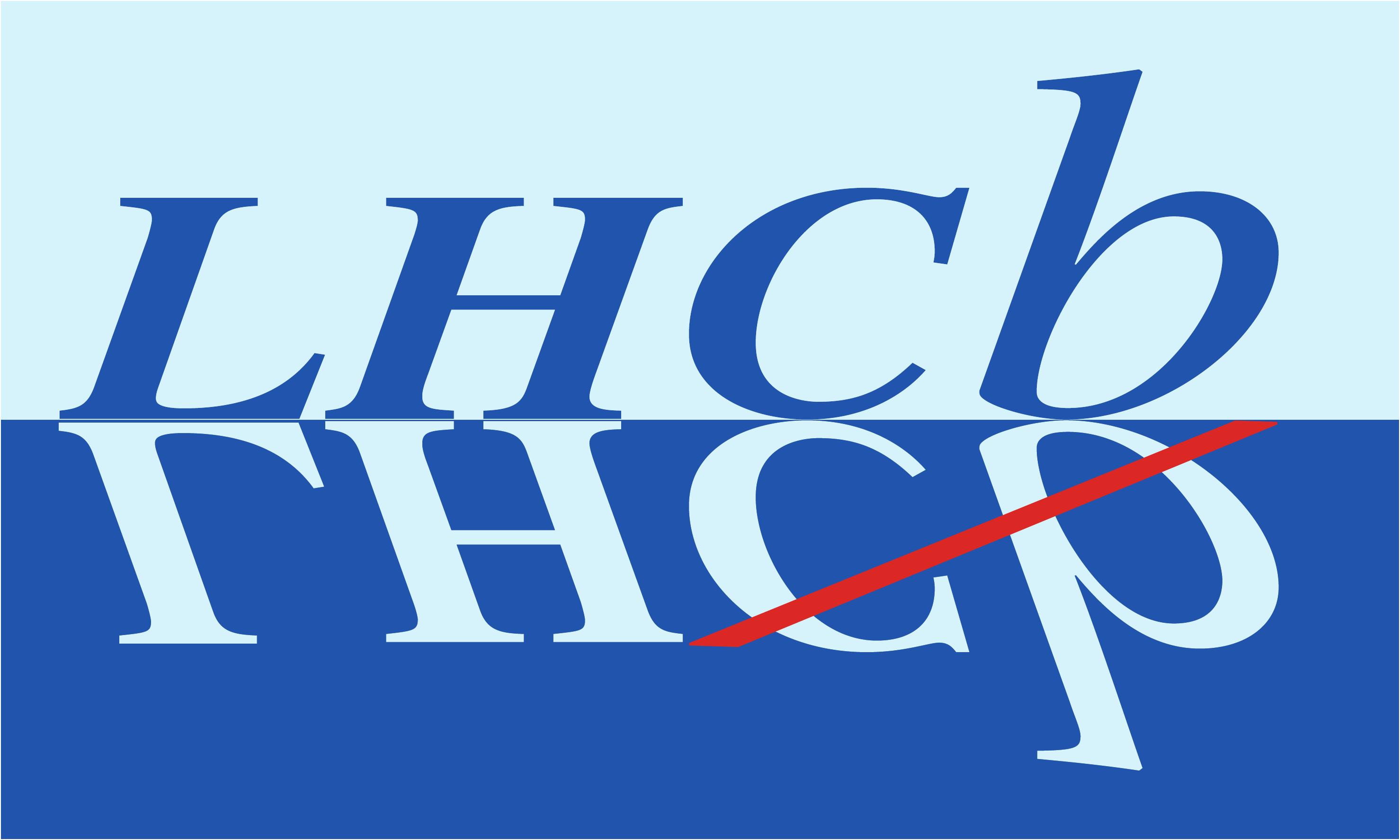}} & &%
\\
& & CERN-EP-2025-150 \\  % ID 
& & LHCb-PAPER-2025-018 \\  % ID 
& & December 8, 2025 \\ % Date - Can also hardwire e.g.: 23 March 2010
& & \\
% not in paper \hline
\end{tabular*}

\vspace*{4.0cm}

% Title --------------------------------------------------
{\normalfont\bfseries\boldmath\huge
\begin{center}
% DO NOT EDIT HERE. Instead edit macro in main.tex to keep metadata correct
\papertitle 
\end{center}
}

\vspace*{2.0cm}

% Authors -------------------------------------------------
\begin{center}
%In the footnote, replace 'paper' by 'Letter' in case of submission to PRL or PLB 
% Edit macro in main.tex to keep metadata correct
\paperauthors\footnote{Authors are listed at the end of this paper.}
\end{center}

\vspace{\fill}

% Abstract -----------------------------------------------
\begin{abstract}
\noindent 
A search for the decay $B^0 \rightarrow \phi \phi$ is made using $pp$ collision data collected with the LHCb detector at centre-of-mass energies of 7, 8 and 13 TeV, corresponding to an integrated luminosity of $9 \, \invfb$. No significant signal is observed, and an upper limit on the branching fraction of $1.3~(1.4)\times 10^{-8}$ at $90 ~(95) \%$ confidence level is set. This result supersedes the previous LHCb study and improves the upper limit by a factor of two. 
\end{abstract}

\vspace*{2.0cm}

\begin{center}
Published in JHEP 12 (2025) 026
\end{center}

\vspace{\fill}

{\footnotesize 
% Edit macro in main.tex to keep metadata correct
\centerline{\copyright~\papercopyright. \href{\paperlicenceurl}{\paperlicence}.}}
\vspace*{2mm}

\end{titlepage}

%%%%%%%%%%%%%%%%%%%%%%%%%%%%%%%%
%%%%%  EOD OF TITLE PAGE  %%%%%%
%%%%%%%%%%%%%%%%%%%%%%%%%%%%%%%%

%  empty page follows the title page ----
\newpage
\setcounter{page}{2}
\mbox{~}

\renewcommand{\thefootnote}{\arabic{footnote}}
\setcounter{footnote}{0}

\cleardoublepage

\pagestyle{plain} % restore page numbers for the main text
\setcounter{page}{1}
\pagenumbering{arabic}

\section{Introduction}
\label{sec:Introduction}
In the Standard Model, the $\Bd \rightarrow \phi \phi$ decay proceeds
via a loop-suppressed annihilation diagram that is subject to both Okubo–Zweig–Iizuka~\cite{OKUBO1963165,Zweig:vol2,iizuka1966meson} and Cabibbo suppression~\cite{Cabibbo:1963yz}. Theoretical predictions for the branching fraction lie in the
range (0.5--5)$\times 10^{-8}$ \cite{Beneke:2006hg,Lu:2005be,Li:2013uaa,BarShalom:2002sv,Zou:2015iwa,Chai:2022ptk,Yan:2022kck,Aleksan:2022mtk}. The large spread in the predictions reflects the
complexity of the perturbative QCD calculations as well as
possible long-range contributions from $\omega-\phi$ mixing,
rescattering and nonfactorizable diagrams that are difficult to estimate. The branching
fraction may be enhanced by up to an order of magnitude in certain new physics
models with additional $\PZ$ bosons \cite{Li:2013uaa} or $R$-parity violating supersymmetry \cite{BarShalom:2002sv}. The first experimental searches for the  $\Bd \rightarrow \phi \phi$  decay were carried out by the CLEO \cite{CLEO:1998hej}, SLD \cite{SLD:1999kau} and
\babar collaborations \cite{BaBar:2008zea}. More recently, the LHCb collaboration set the most
stringent limit to date, $\mathcal{B}(\Bd \to \phi \phi) < 2.7 \times 10^{-8}$ at $90 \%$ confidence
level~\cite{LHCb-PAPER-2019-019}, using data collected between 2011 and 2016. 

 In this paper, a dedicated search is made for the $\Bd \rightarrow
 \phi \phi$ decay mode,\footnote{Charge conjugation is implied throughout.} with the subsequent decay $\phi \rightarrow \Kp \Km$, where $\phi$ implies the $\phi(1020)$ throughout. The abundant signal of the corresponding $\Bs$ mode provides an ideal control channel and normalisation mode, with branching fraction $\mathcal{B}(\Bs \rightarrow \phi \phi) = (1.84 \pm 0.05 \, \stat \pm 0.07  \, \syst \pm 0.011 \, (f_s/f_d) \, \pm 0.12 \, \textrm{(norm)}) \times 10^{-5}$ ~\cite{LHCb-PAPER-2015-028}.  
 This study, which supersedes the previous LHCb result~\cite{LHCb-PAPER-2019-019}, uses the full dataset collected by the LHCb
 collaboration in proton-proton ($\proton\proton$) collisions
 between 2011 and 2018, which corresponds to an integrated luminosity
 of 3\invfb collected at a centre-of-mass energies of $\sqrt{s}=7$ and $8 \tev$ (Run 1) and 
 6\invfb collected at $\sqrt{s}=13 \tev$ (Run 2). The dataset used is a factor of two larger than that used
 in the previous LHCb study \cite{LHCb-PAPER-2019-019} which analysed only the data collected up to 2016. In addition, a more detailed understanding of background sources has allowed for significant reduction of the background, which further improves the sensitivity by a factor of 1.8. To avoid experimenter bias, the invariant-mass
 region around the $\Bd$ mass was not examined until the full analysis procedure had been finalised. 

%%%%%%%%%%%%%%%%%%%%%%%%%%%%%%%%%%%%%%%%%%%%%%%%%%%%%%%%%%%%%%%%%%%%%%%%%%%%%%%%%%%%%%%%%%%%%%%%%
\section{Detector and simulation}
\label{sec:Detector}
The \lhcb detector~\cite{LHCb-DP-2008-001,LHCb-DP-2014-002} is a
single-arm forward spectrometer that covers the~\mbox{pseudorapidity}
range $2<\eta <5$, designed for the study of particles containing
\bquark or \cquark quarks. The detector used for this analysis includes a high-precision tracking
system consisting of a silicon-strip vertex detector
surrounding the $\proton\proton$~interaction region, a large-area
silicon-strip detector located upstream of a dipole magnet with 
a bending power of approximately $4$~T\,m, and three stations of 
silicon-strip detectors and straw drift tubes  placed downstream of the~magnet. 
The~tracking system provides a measurement of the momentum, \ptot,  
of charged particles with a relative uncertainty that varies from~0.5\% at low momentum  
to~1.0\% at $p=200\gevc$.  Large samples of $\decay{\Bu}{\jpsi\Kp}$ and inclusive $\decay{\jpsi}{\mumu}$~decays, collected concurrently with
the~dataset used in this analysis, are used to calibrate the momentum
scale of the spectrometer~\cite{LHCb-DP-2023-003}. The relative uncertainty achieved on the momentum scale is $3 \times 10^{-4}$. 

Charged hadrons are distinguished using information from two ring\nobreakdash-imaging 
Cherenkov\,(RICH) detectors. In addition, photons, electrons and hadrons are 
identified by a calorimeter system consisting of scintillating-pad and preshower 
detectors, an electromagnetic and a hadronic calorimeter. Muons are
identified by a system composed of alternating  layers of iron and multiwire proportional chambers.

The online event selection is performed by a~trigger,  which consists
of a~hardware stage followed by a~two\nobreakdash-level software
stage~\cite{LHCb-DP-2012-004,LHCb-DP-2019-001}. At the hardware
trigger stage, events are required to have a muon with high transverse momentum or a
hadron, photon or electron with high transverse energy in the
calorimeters. For hadrons,  the transverse energy threshold is
3.5\gev. The software trigger requires a two-, three- or four-track
secondary vertex with a significant displacement from any primary $pp$
interaction vertex (PV). At least one charged particle must have a
transverse momentum $\pt > 1.6\gevc$ and be inconsistent with
originating from a PV. A multivariate
algorithm~\cite{BBDT,LHCb-PROC-2015-018} is used for the identification of secondary
vertices consistent with the decay of a \bquark hadron. In the offline
selection, trigger signals are associated with reconstructed
particles. Selection requirements can therefore be made on the trigger
selection itself and on whether the decision was due to the signal candidate (TOS) or other
particles produced in the $pp$ collision (TIS). 

Simulation is used to determine selection requirements and to describe the
invariant-mass distribution of the signal candidates. To accurately model the detector
resolution, $pp$ collisions are generated using
\pythia~\cite{Sjostrand:2007gs,*Sjostrand:2006za}
with a specific \lhcb configuration~\cite{LHCb-PROC-2010-056}. The decays
of unstable particles
are described by \evtgen~\cite{Lange:2001uf}, in which final-state
radiation is generated using \photos~\cite{davidson2015photos}.
The interaction of the generated particles with the detector, and its
response, are implemented using the \geant
toolkit~\cite{Allison:2006ve, *Agostinelli:2002hh} as described in Ref.~\cite{LHCb-PROC-2011-006}. The full detector simulation reproduces the mass resolution observed in the data for \bquark-hadron decays at the level of $5\%$ or better. Several sources of background are also studied using the~{\sc{RapidSim}} fast-simulation package~\cite{Cowan:2016tnm}. 

% \section{Interaction MVA}
\section{Classifier for controlling kaon interactions}
\label{sec:interactions}
Assuming a $\Bd \to \phi \phi$ branching fraction at the $10^{-8}$ level, the $\Bs \rightarrow \phi \phi$ decay proceeds with a rate approximately three orders of magnitude larger. This, combined with a mass resolution of $15 \mevcc$ and $m(\Bs) - m(\Bd) =87.22 \pm 0.12 \mevcc$ \cite{PDG2024} means it is important to control the tail of the $\Bs$ mass distribution within the $\Bd$ signal region. Simulation studies show that the tail is dominated by candidates where one or more kaons either decay in flight or interact hadronically within the tracking system. Suppression of decays-in-flight is possible by removing kaon candidates with associated activity in the muon detector. However, this requirement alone is not sufficient, as it does not reject either hadronic interactions or decays-in-flight outside the muon system acceptance.

\begin{figure}
    \centering
    \includegraphics[width=0.7\linewidth]{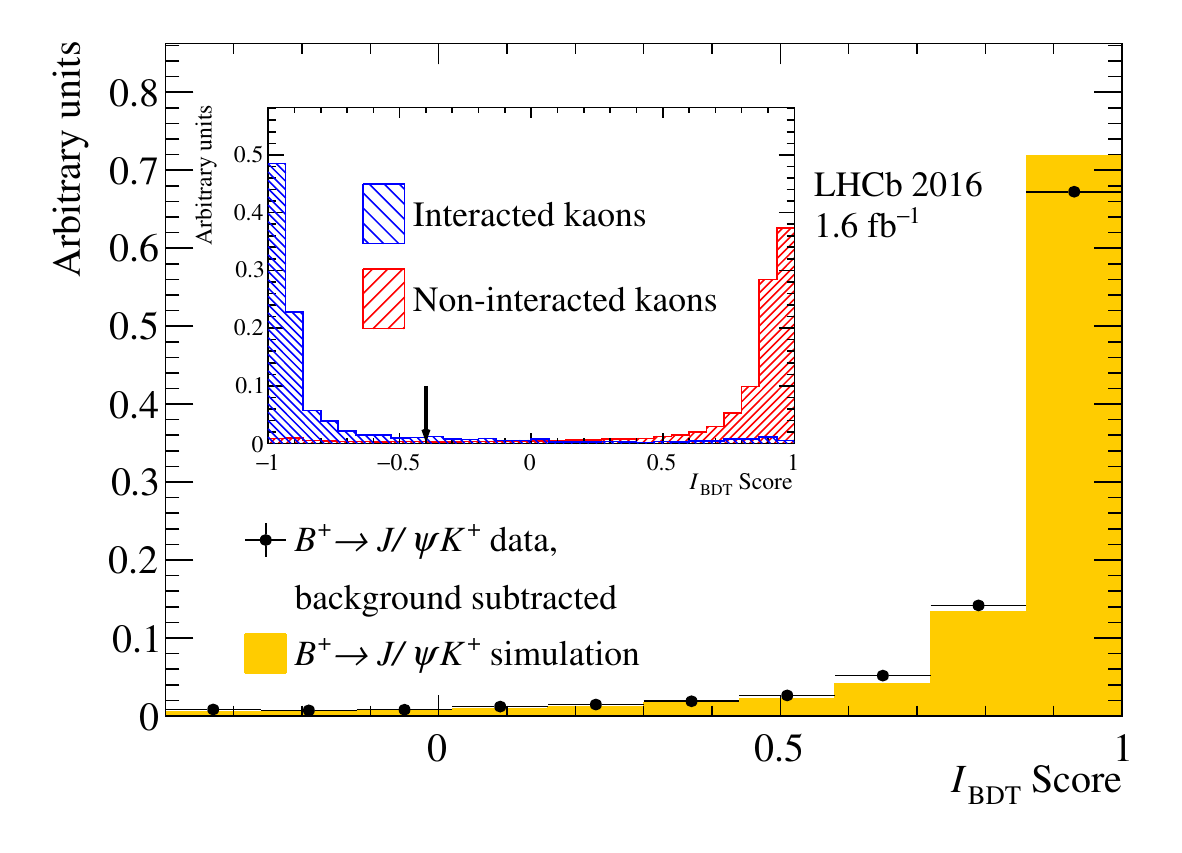}
    \caption{Distribution of the $I_{\textrm{BDT}}$ classifier score for muons from the $\Bp \rightarrow \jpsi \Kp$ decay in simulation and the LHCb 2016 data sample as an example. In the data, the combinatorial background has been subtracted using the technique described in Ref.~\cite{Dembinski:2021kim}. The statistical uncertainties on the data points are plotted but not visible. The inset figure shows the classifier response for interacted and non-interacted kaons in the $\Bs \rightarrow \phi\phi$ simulation samples used for training. The arrow on the inset figure indicates the imposed requirement. }
    \label{fig:bdtg_agreement}
\end{figure}

\begin{figure}
    \centering
    \includegraphics[width=0.7\linewidth]{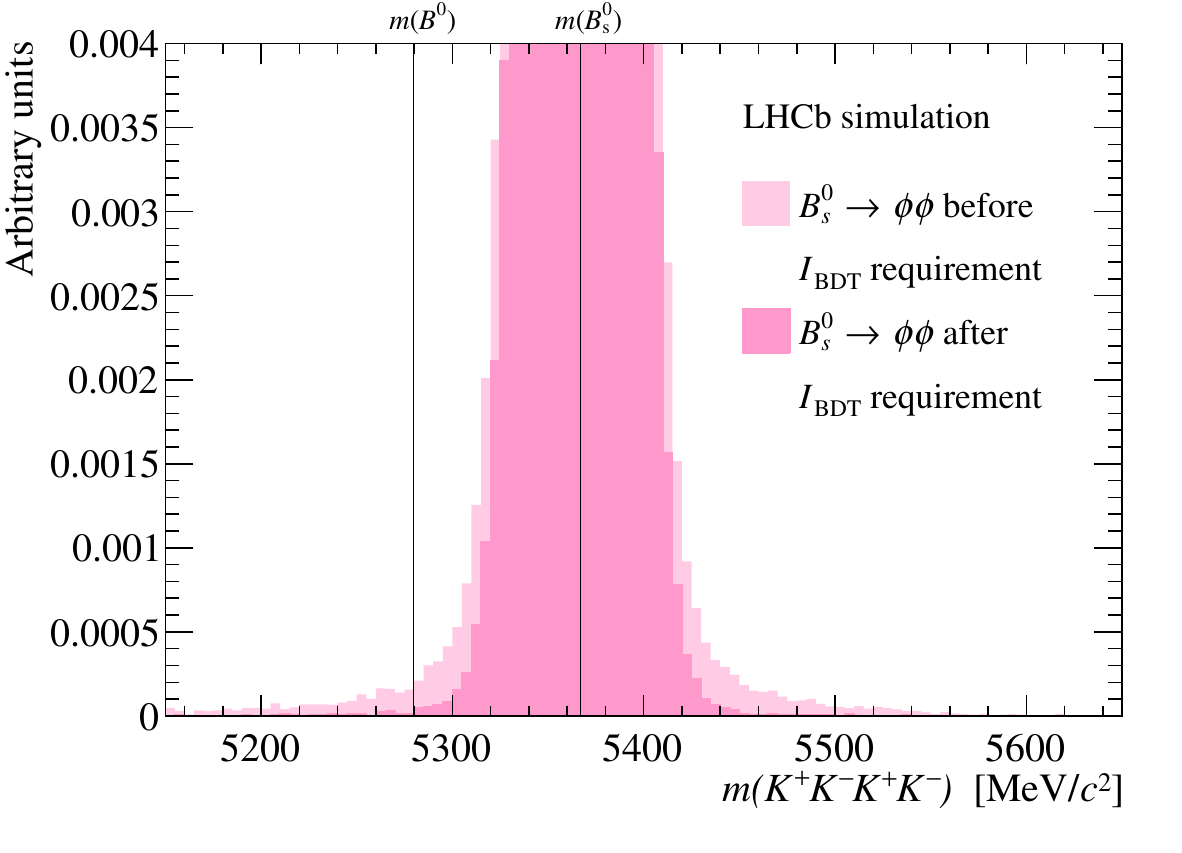}

    \caption{Four-kaon invariant-mass distribution of simulated $\Bs \to \phi\phi$ events shown before and after the application of the $I_{\textrm{BDT}}$ requirement. The histograms are scaled such that the distribution before application of the classifier has unit area. The known $\Bd$ and $\Bs$ masses are indicated with solid black lines~\cite{PDG2024}. The full selection described in Section~\ref{sec:selection} has been applied.}
    
    \label{fig:bdtg_before_after}
\end{figure}

A well-established technique to discriminate between muons and hadrons that decay in flight or interact is to identify breakpoints or discontinuities in the trajectory of the particle within the tracking system \cite{Astier:1999rs,Belle:2023dyc,CMS:2024qyo}. This approach is adopted here with the novelty that a multivariate technique is used to separate kaons with and without interactions. Using simulation, a gradient-boosted decision tree \cite{Breiman,AdaBoost} from the {\sc{TMVA}} toolkit \cite{Hocker:2007ht,*TMVA4} is used to train a classifier, $I_{\textrm{BDT}}$. The truth information available in the simulation allows for identification of kaons that have interacted within the tracking system volume. As input, the $I_{\textrm{BDT}}$ classifier uses discriminating variables related to the track fit quality, the matching of the track segments upstream and downstream of the spectrometer magnet, the number of hits on the track and the track kinematics. These variables are chosen as they show good agreement between data and simulation for muons from the $\Bp \rightarrow \jpsi (\rightarrow \mumu) \Kp$ decay chain which acts as a control channel. As shown in Fig.~\ref{fig:bdtg_agreement}, good separation between kaons with and without interactions is achieved. The $I_{\textrm{BDT}}$ score shows reasonable agreement between data and simulation for the muons from the $\Bp \rightarrow \jpsi \Kp$ decay which are used as a proxy for non-interacted kaons. Requiring $I_{\textrm{BDT}} > -0.4$ rejects $89\%$ of interacted kaons with an efficiency of $95\%$ for non-interacted kaons.\footnote{Here, `interacted' refers to kaons that have either interacted hadronically or decayed in flight.} The efficiency of the chosen requirement agrees between data and simulation at the level of 1\%. The effect of the $I_{\textrm{BDT}}$ requirement is shown for simulated $\Bs \to \phi \phi$ events in Fig.~\ref{fig:bdtg_before_after}. It can be seen that after the $I_{\textrm{BDT}}$ requirement the tails from $\Bs \rightarrow \phi \phi$ are significantly reduced. This is an appreciable improvement compared to the previous analysis \cite{LHCb-PAPER-2019-019} where these tails limited the sensitivity to a potential $\Bd \to \phi \phi$ signal. 

\section{Selection}
\label{sec:selection}
The offline selection consists of a loose preselection, including the rejection of poorly reconstructed candidates that have interacted in the spectrometer. This is followed by requirements on two multivariate algorithms that efficiently reduce the combinatorial background while maintaining high signal efficiency.

The preselection requires four charged particles that have a good quality track fit~\cite{LHCb-DP-2014-002}, $\pt > 400 \, \mevc$ and that are well-identified as kaons by the RICH detectors. Kaons that interact within the tracking system are rejected by the $I_{\textrm{BDT}}$ requirement, described in Section~\ref{sec:interactions}, and also by requiring no associated activity in the muon stations. Fake tracks created by the reconstruction from random combinations of hits in the detector are suppressed using a neural network trained to discriminate between these and genuine tracks associated to particles~\cite{DeCian:2255039}. Oppositely charged kaon pairs that form a common vertex are combined to form $\phi$ candidates. Background from \bquark-hadron decays with misidentified particles, such as $\Bd \rightarrow \phi \Kstarz$ and $\Lb \rightarrow \phi p \Km$, is strongly reduced by requiring the invariant mass of both $\Kp\Km$ pairs to be within $\pm 15 \mevcc$ of the known mass of the $\phi(1020)$ resonance~\cite{PDG2024}.

Selected $\phi$ candidates are combined to form $\B$ candidates. To reduce combinatorial background, the product of the transverse momenta of the $\phi$ mesons is required to be greater than $1.2 \gev^2\!/c^2$ and that the reconstructed $\B$~decay time must exceed $0.3\ps$.  A vertex fit is performed on each candidate \cite{Hulsbergen:2005pu} requiring that the four tracks originate from a common point, and that the $\B$ candidate originates from the associated PV, chosen as the PV that best matches the flight direction of the $\B$ candidate. This fit is required to converge and a loose requirement is made on the $\chi^2$ of this fit. To ensure good separation between $\Bs$ and $B^0$ decays to the $\phi\phi$ final state, the estimated uncertainty on the mass of the $\B$ candidate is required to be less than $20 \mevcc$. Background from the $\Bd \rightarrow \phi K^*$ decay is suppressed by taking the kaon with the highest probability of being misidentified as a pion and assigning it the pion mass. If either $m(\Kp\pim)$ is within $\pm70 \mevcc$ of the known $\Kstarz$ mass \cite{PDG2024} or $m(\Kp\Kp\Kp\pim)$ is within $\pm30 \mevcc$ of the known $\Bd$ mass, tighter particle-identification requirements are applied. 
 
After applying these criteria, although the mass peak for $\Bs$ signal is clearly visible, the combinatorial background due to random combinations of $\phi$ mesons is too large to give sufficient sensitivity for the $\Bd$ search, and thus needs to be further reduced. Detached $\phi$ mesons are abundantly produced in $\Ds$ meson decays. Since \mbox{$\mathcal{B}(\Ds \rightarrow \phi X) = (15.7 \pm 1.0)\%$}~\cite{PDG2024}, this decay is expected to be an important background source. Combining a candidate $\phi$ meson used to form a selected \BdorBs candidate with a high-$\pt$ pion that is displaced from all PVs, a clear peak is seen at the $\Ds$ mass, confirming this hypothesis. Candidate $\Ds$ mesons selected in this way are used to train a gradient-boosted decision tree classifier \cite{Breiman,AdaBoost} from the {\sc{TMVA}} toolkit \cite{Hocker:2007ht,*TMVA4}, referred to as the DBDT. This classifier is trained on data alone and is designed to select generic three-track combinations consistent with a $\Ds$ decay. To determine the training weights a fit is made to the $K^+K^-\pi^+$ invariant-mass distribution and the \sPlot technique \cite{Pivk:2004ty} is used to separate signal and background. The classifier takes as input information related to the vertex fit $\chi^2$, displacement from the PV and the pseudorapidity of the $\Ds$ candidate.  By making a requirement on DBDT and selecting the region around the known $\Ds$ mass it is possible to cleanly identify and tag 90\% of the background originating from a $\Ds\rightarrow \phi \pi^+$ decay, where the pion is reconstructed. The DBDT classifier also allows for the identification of $\phi \mu^+ \nu_{\mu}$ and $\phi e^+ \nu_{e}$ decays. In this case, since the neutrino is not reconstructed, only a loose requirement can be made on the $\phi l^+$ invariant mass. However, this is compensated by the lower combinatorial background for displaced high-$\pt$ electrons or muons and this enables selection of these topologies with a low misidentification rate. In total, around $10 \%$ of the preselected candidates are identified as being from a $\Ds$ decay. These candidates are referred to as $\Ds$-tag background and are rejected with negligible loss in signal efficiency. The three decay modes considered account for around $50\%$ of the inclusive branching fraction $\mathcal{B}(\Ds \rightarrow \phi X)$. The remainder of the branching fraction is dominated by modes with multiple charged or neutral pions, which cannot be cleanly identified.  Given the $\Ds$-tag only attempts to identify $50\%$ of $\Ds \rightarrow \phi X$ decays and that the acceptance and reconstruction efficiency for the additional particle is $\sim 60 \%$, a sizeable fraction of the remaining background is due to untagged candidates with this topology. Consequently, the $\Ds$-tag sample is used, together with simulated $\Bs \rightarrow \phi \phi$ decays, to train a multilayer perceptron (MLP) from the {\sc{TMVA}} toolkit \cite{Hocker:2007ht}. This identifies $\Ds \to \phi X$ background candidates where no tag was possible. In the MLP, variables related to the $B$ candidate isolation, the vertex fit $\chi^2$ and the $\chisqip$ are used to discriminate between signal and background.\footnote{The $\chisqip$ is defined as the difference in the vertex-fit $\chi^2$ of a given PV reconstructed with and without the particle under consideration.} To allow for differences in data-taking conditions, separate classifiers are trained for the Run~1 and Run~2 datasets. The chosen requirement on the MLP output rejects $99.0\%$ of the $\Ds$-tag sample whilst keeping $97.5\%$ of simulated signal candidates.

\begin{figure}
    \centering
    
    \includegraphics[width=0.49\linewidth]{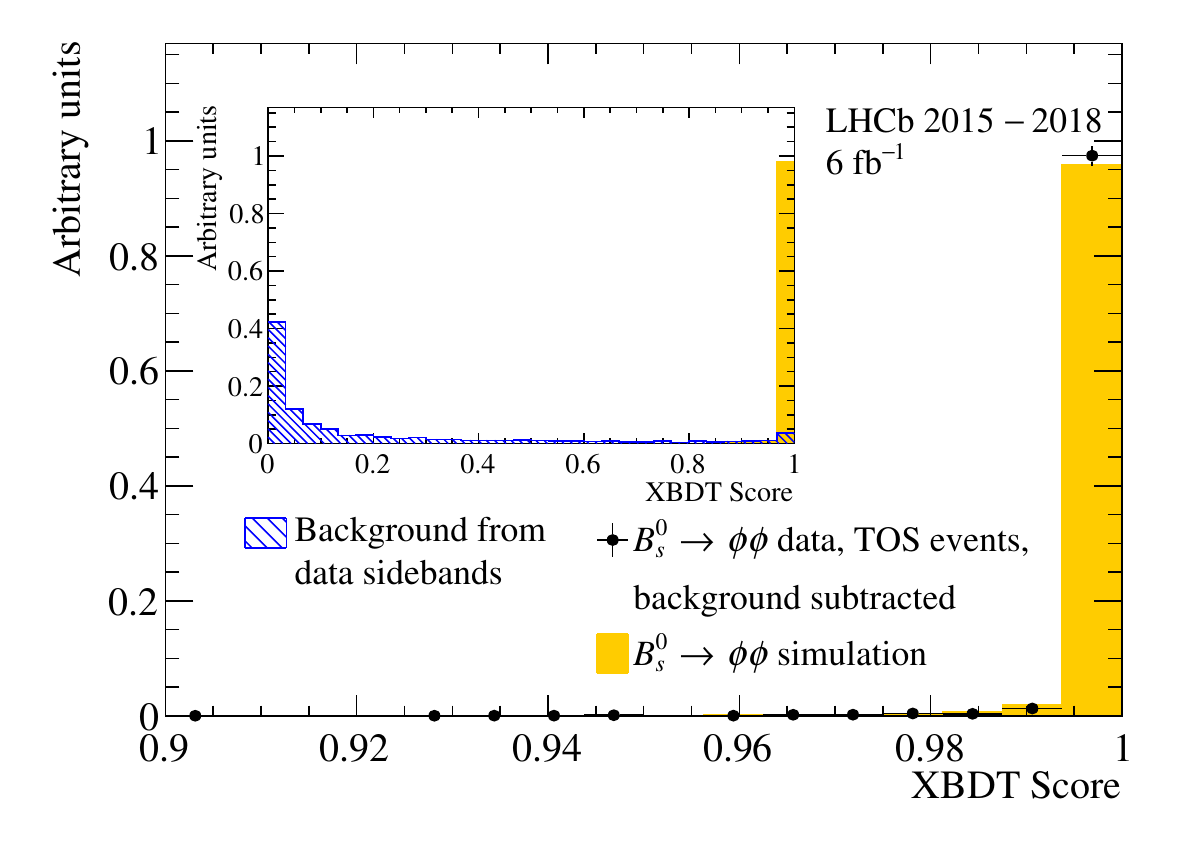}
    \includegraphics[width=0.49\linewidth]{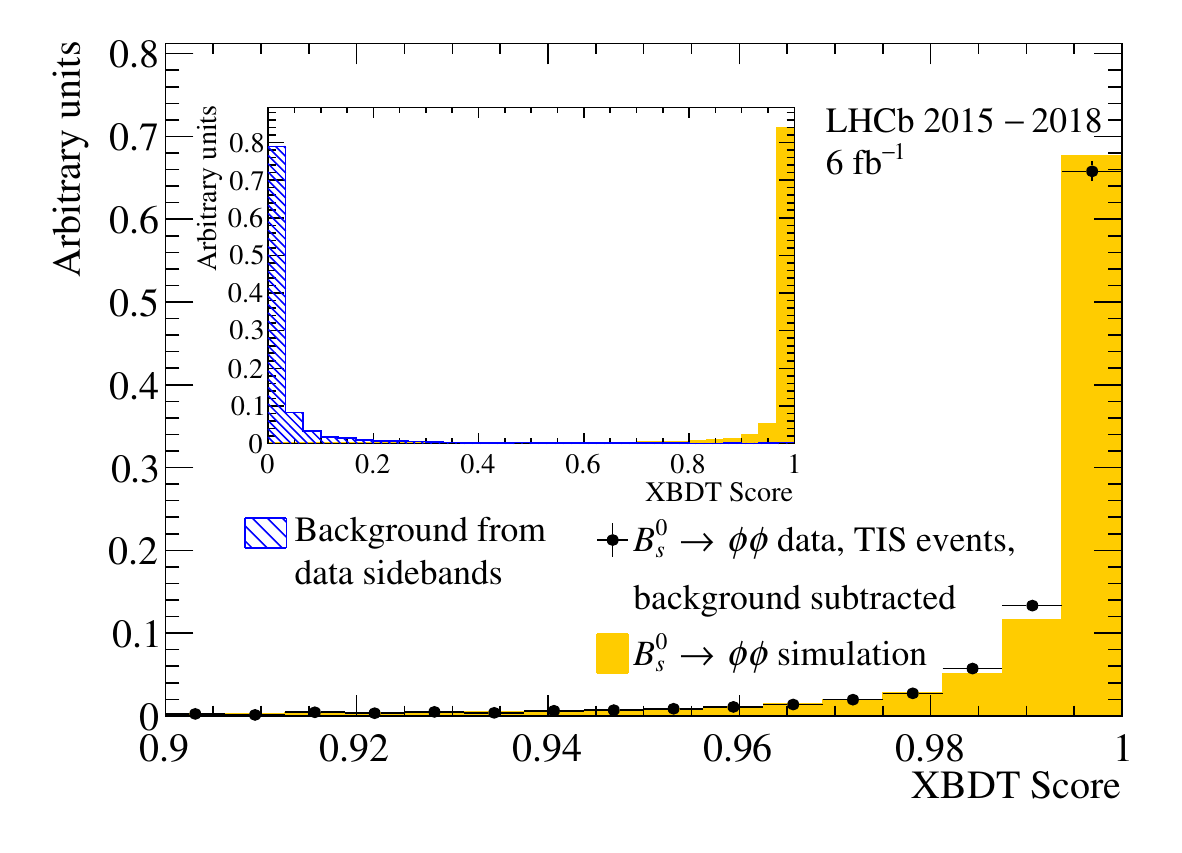}
    \caption{Response of the XBDT classifier for the Run~2 (left) TOS and (right) TIS categories for the $\Bs \rightarrow \phi\phi$ component in data (black points) and simulation (solid yellow). The response of the classifier on the background dataset used for training is shown in hatched blue in the inset plot.  Combinatorial background in the data has been subtracted using the approach described in Ref.~\cite{Dembinski:2021kim}.}
    
    \label{fig:XBDT}
\end{figure}

The final step of the selection procedure aims to discriminate between signal and generic combinatorial background. For this purpose, a gradient-boosted decision tree classifier (XBDT) is trained using the {\sc{XGBoost}} package \cite{Chen:2016:XST:2939672.2939785}. The training is performed separately for the TIS and TOS trigger categories, as they have different kinematic distributions and background levels, and also separately for the Run~1 and Run~2 data-taking periods. The classifier is trained using the $\Bs \rightarrow \phi\phi$ simulation samples as a proxy for the signal, and the $\B$ mass sidebands of the data as a proxy for the background. The sidebands are defined as $m(K^+K^-K^+K^-) < 5100$~\mevcc{} or $ m(K^+K^-K^+K^-) > 5650$~\mevcc{}. Only candidates passing a loose requirement on the MLP classifier score are used for training. This requirement was chosen to have sufficient background candidates for training while removing candidates that the MLP can easily classify as background. Variables related to the $\BdorBs$ candidate kinematics, displacement, activity in the cone around the candidate \cite{LHCb-PAPER-2016-054}, the isolation of the secondary vertex, the vertex fit $\chi^2$, together with track fit $\chi^2$ and particle-identification information of the kaons, are used in the training. All the variables used provide good separation between signal and background and agree well between data and simulation. As an example, Fig.~\ref{fig:XBDT} shows the classifier performance for the Run~2 TOS and TIS categories. To optimise the XBDT selection requirement for each category, the multicategory generalisation of the figure-of-merit described in Refs.~\cite{Punzi:2020fsv,Punzi:2003bu} is used, 
\begin{equation}
    \text{FOM} = \frac{A^2/B}{a/2 + A^{3/2}/B},
\end{equation}
with,
\begin{equation}
    A = \sum_i \frac{\epsilon^2_i}{b_i}\quad,\quad B = \sum_i \frac{\epsilon^3_i}{b_i^2},
\end{equation}
where $\epsilon_i$ is the efficiency of the selection on signal in category $i$ in $\{\text{TIS},\text{TOS}\}$, $b_i$ is the estimated background level in the invariant-mass region around the $\Bd$ mass in category $i$ with the parameter $a$ set to $3$, corresponding to a $3\sigma$ evidence. The chosen working points of the XBDT classifiers select signal with an overall efficiency of 92.7\% and 82.6\% in Run~1 and Run~2, respectively.

\section{Invariant-mass fit model}
\label{sec:massmodel}
The simulation is used to develop invariant-mass models for the signal and background components for the Run~1 and Run~2 datasets. An important consideration in these studies is to ensure that the more abundant $\Bs \rightarrow \phi \phi$ decay is well described. Using the simulation, it is found that this component can be modelled by the sum of two Student's t-distributions~\cite{studentt} and a wide Gaussian component, with a common mean. The Gaussian component accounts for the remaining tail from interacted kaons.  When fitting the data, the common mean and the $n$-parameter of the narrower Student's t-distribution are left free for the $\Bs$ component. The $n$-parameter of the wider Student's t-distribution together with the width from the Gaussian function are constrained with Gaussian priors based on simulation fits, while the remaining parameters are fixed to the simulation values. The same model is used for the possible $\Bd \to \phi \phi$ component. In that case, the mean is constrained using the known $\Bd-\Bs$ mass splitting \cite{PDG2024}, while the resolution of each of the Student's t-distributions is constrained via a Gaussian prior centered on the resolution of the $\Bs$ component, with a scale factor of $0.996 \pm 0.003$ determined from the simulation.

Several background contributions are included in the invariant-mass model. The dominant background source is combinatorial and is modelled using an exponential function. As an alternative, a power law model is considered and found to give consistent results. Background from \mbox{$\Lb \rightarrow \phi p \Km$} and \mbox{$\Xibz \rightarrow \phi p \Km$} decays, where the proton is misidentified as a kaon, gives a broad shoulder to the right of the $\Bs$ mass peak. Using the simulation, the shape of these components is found to be well described by Crystal Ball functions \cite{Skwarnicki:1986xj}. The yield of these components is estimated using data enriched in misidentified particles, obtained by selecting events that do not satisfy the particle-identification criteria used to separate kaons from protons. The number of rejected background candidates is estimated from data that fail the particle-identification criteria used to select the signal and then multiplied by the pass-fail factor obtained from a simulated sample of $\Lb \rightarrow \phi p \Km$ decays. This is done separately for Run~1 and Run~2 and the results applied as a Gaussian constraint on the yield of this component in data. The remaining background from $\Bd \rightarrow \phi \Kstar(892)^0$ decays, with a misidentified pion, is modelled with a Crystal Ball function using simulation. The yield of this component is constrained relative to the $\Bs \rightarrow \phi \phi$ signal using the relative efficiency from the simulation and the known branching fractions. The yields relative to the $\Bs \rightarrow \phi \phi$ signal are constrained to $(24\pm4)\times10^{-4}$ and $(22\pm2)\times10^{-4}$ in Run 1 and Run 2, respectively. Finally, partially reconstructed background from \bquark-hadron decays to charged and neutral $K_1(1270)$ and $K_2^{*}(1430)$ intermediate states is studied and modelled with a kernel estimation probability density function~\cite{Cranmer:2000du} using samples generated with {\sc{RapidSim}}~\cite{Cowan:2016tnm}. In the data fit the yield of this component is constrained using the known branching fractions and relative efficiency from the simulation. The yield relative to the $\Bs \rightarrow \phi \phi$ signal is constrained to $(12\pm2)\times10^{-4}$.
  
The adopted fit model has been extensively validated using pseudoexperiments. In particular, it has been verified that the model is unbiased and does not generate a spurious signal following the procedure outlined in Ref.~\cite{ATL-PHYS-PUB-2020-028}.  

\section{Results and conclusions}
\label{sec:results}
\begin{figure}[tb!]
    \centering
    \includegraphics[width=\linewidth]{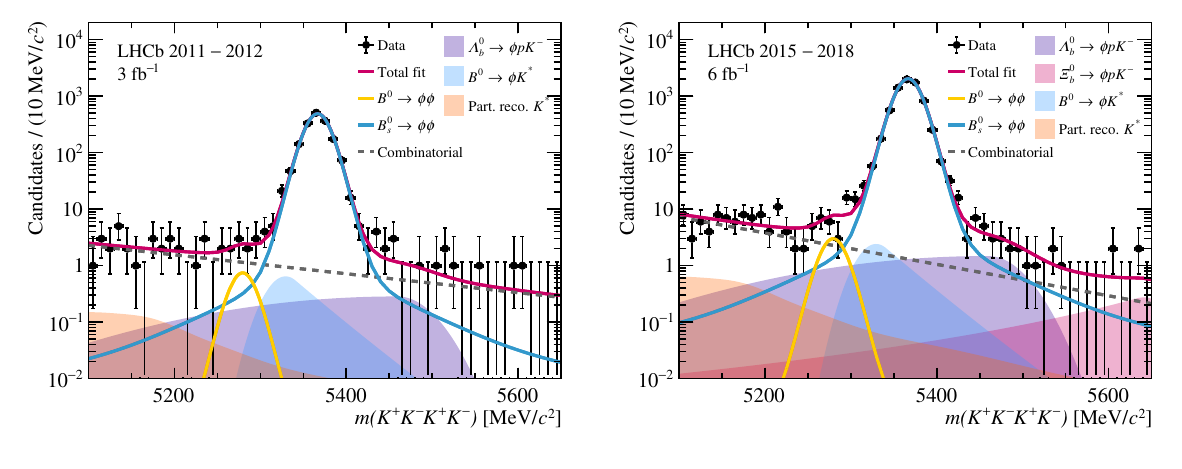}
    
    \caption{Four-kaon invariant-mass distribution for the (left) Run~1 and (right) Run~2 datasets. The fit result is also shown. }
    \label{fig:massfit}
\end{figure}

To determine the $\Bd \rightarrow \phi \phi$ signal yield, an extended unbinned maximum-likelihood fit with the model described in Section~\ref{sec:massmodel} is performed to the Run 1 and Run 2 datasets simultaneously. The yield of $\Bs \to \phi \phi$ events is free to vary in the fit. The yields for the background components are calculated separately for Run~1 and Run~2 while all shapes are shared except the combinatorial background. The results of this fit are shown in Fig.~\ref{fig:massfit}. The fit yields $2.6\pm1.5$ and $10.5\pm6.3$ signal decays for the Run~1 and Run~2 datasets, respectively, where the uncertainties quoted are statistical only.
The results are consistent with the background-only hypothesis at the $1.9\sigma$ level. Hence, limits are set using the $\textrm{CL}_{s}$ method \cite{CLs} with a one-sided test statistic, evaluated using pseudoexperiments. Systematic effects from modelling the signal and background components and knowledge of external branching fractions are included in the limit via Gaussian constraints. The largest source of systematic uncertainty comes from the $\Bs \to \phi \phi$ branching fraction which has a relative uncertainty of 7.7\%. The uncertainty on the selection efficiency ratio gives a relative uncertainty of 0.5\% on the central value due to the limited size of simulation samples. Since the limit on the $\Bd \rightarrow \phi \phi$ branching fraction is calculated with respect to the $\Bs\rightarrow \phi \phi$ decay with the same final state, several sources of systematic uncertainty cancel in the ratio. This includes systematic uncertainties due to tracking, particle identification and trigger efficiencies. The measured $\Bd \rightarrow \phi \phi$ yield, $N_{\Bd}$, is converted into a branching fraction using
\begin{equation*}
  \mathcal{B}(\Bd \rightarrow \phi \phi) = \frac{N_{\Bd}}{N_{\Bs}} \cdot \frac{\epsilon_{\Bs\rightarrow \phi \phi}}{\epsilon_{\Bd\rightarrow \phi \phi}}  \cdot  \mathcal{B}(\Bs \rightarrow \phi \phi) \cdot f_s/f_d,
\end{equation*}
where $N_{\Bs}$ is the $\Bs \rightarrow \phi \phi$ yield, $f_s/f_d$ is the ratio of the fragmentation fractions and $\mathcal{B}(\Bs \rightarrow \phi \phi) = (1.84 \pm 0.05 \, \stat \pm 0.07  \, \syst \pm 0.011 \, (f_s/f_d) \, \pm 0.12 \, \textrm{(norm)}) \times 10^{-5}$ \cite{LHCb-PAPER-2015-028}. The calculation takes into account the correlation between the external LHCb measurements, including those reported in~\cite{LHCb-PAPER-2015-028}. In particular, the contribution of $f_s/f_d = 0.259 \pm 0.015$ cancels in the calculation given the decay $\Bd \rightarrow \phi \Kstarz$ was used for normalisation in Ref.~\cite{LHCb-PAPER-2015-028} where the $\Bs \to \phi\phi$ branching fraction is measured, giving $\mathcal{B}(\Bs \rightarrow \phi \phi)\cdot(f_s/f_d) = (477\pm37) \times 10^{-8}$. The ratio of efficiencies, $\epsilon_{\Bd\rightarrow \phi \phi}/\epsilon_{\Bs\rightarrow \phi \phi}$, is evaluated using the simulation to be $1.006 \pm 0.006$ for Run~1 and $0.971 \pm 0.003$ for Run~2, where the uncertainties are statistical only.

\begin{figure}[tb!]
    \centering
    \includegraphics[width=0.7\linewidth]{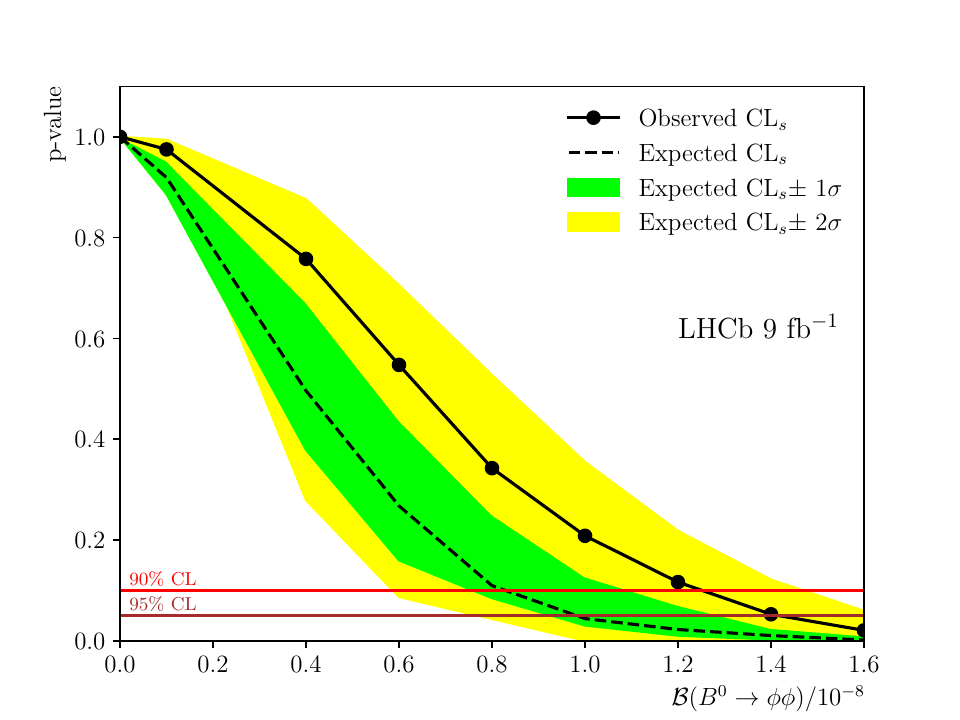}
    \caption{Results of the CL$_s$ method for setting an upper limit on the $\Bd\to\phi\phi$ branching fraction.}
    \label{fig:CLs}
\end{figure}

\begin{figure}[htb!]
    \centering
    \includegraphics[width=0.5
    \linewidth]{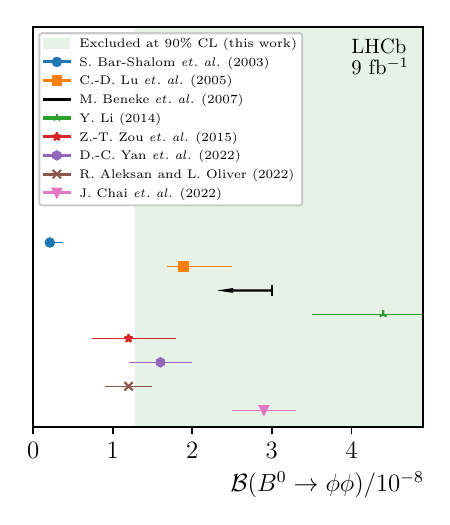}
    \caption{Limit set on $\mathcal{B}(\Bd \rightarrow \phi \phi)$ in this analysis at 90\% confidence level compared to theoretical predictions. The results presented in this paper are represented by the shaded region.}
    \label{fig:predictions}
\end{figure}

The result of the CL$_s$ calculation is shown in Fig.~\ref{fig:CLs}. The limits obtained are
\begin{equation*}
\mathcal{B}(\Bd \rightarrow \phi \phi) < 1.3~(1.4) \times 10^{-8} \, \, \textrm{at 90\% (95\%) confidence level}.
\end{equation*}
The obtained limit is found to be robust to changes in the signal and background modelling in the fit. For example, changing the combinatorial background model, the constraints applied to the signal line shape parameters, the modelling of the $\Lb \rightarrow \phi p \Km$ component and the partially reconstructed component all have no impact on the result. Additionally, relaxing the constraint on the width of the shallow Gaussian component of the signal model has a negligible impact on the result. The result favours the lower end of the range of current predictions for the branching fraction (Fig.~\ref{fig:predictions}) and strongly constrains possible enhancements due to physics beyond the Standard Model. It supersedes the previous LHCb limit, presented in Ref.~\cite{LHCb-PAPER-2019-019}, improving on it by a factor of two. Further improvements to this limit or evidence for this decay mode will be possible using the larger data set that is being collected by the upgraded LHCb experiment. Those studies will benefit from the improved understanding of the experimental backgrounds achieved in this analysis.

\section*{Acknowledgements}
%
% These Acknowledgements valid from 12/09/25
%
\noindent We express our gratitude to our colleagues in the CERN
accelerator departments for the excellent performance of the LHC. We
thank the technical and administrative staff at the LHCb
institutes.
We acknowledge support from CERN and from the national agencies:
ARC (Australia);
CAPES, CNPq, FAPERJ and FINEP (Brazil); 
MOST and NSFC (China); 
CNRS/IN2P3 (France); 
BMFTR, DFG and MPG (Germany);
INFN (Italy); 
NWO (Netherlands); 
MNiSW and NCN (Poland); 
MCID/IFA (Romania); 
%MSHE (Russia); 
MICIU and AEI (Spain);
SNSF and SER (Switzerland); 
NASU (Ukraine); 
STFC (United Kingdom); 
DOE NP and NSF (USA).
%%%%%%%%%%%%%%%%%%%%%%%%%%%%%%%%%%%%%%%%%%%%%
We acknowledge the computing resources that are provided by ARDC (Australia), 
CBPF (Brazil),
CERN, 
IHEP and LZU (China),
IN2P3 (France), 
KIT and DESY (Germany), 
INFN (Italy), 
SURF (Netherlands),
Polish WLCG (Poland),
IFIN-HH (Romania), 
%RRCKI and Yandex LLC (Russia), 
PIC (Spain), CSCS (Switzerland), 
and GridPP (United Kingdom).
%%%%%%%%%%%%%%%%%%%%%%%%%%%%%%%%%%%%%%%%%%
We are indebted to the communities behind the multiple open-source
software packages on which we depend.
%%%%%%%%%%%%%%%%%%%%%%%%%%%%%%%%%%%%%%%%%%
Individual groups or members have received support from
%ARC and ARDC (Australia); % moved to national 16/01
Key Research Program of Frontier Sciences of CAS, CAS PIFI, CAS CCEPP, 
Fundamental Research Funds for the Central Universities,  and Sci.\ \& Tech.\ Program of Guangzhou (China);
Minciencias (Colombia);
EPLANET, Marie Sk\l{}odowska-Curie Actions, ERC and NextGenerationEU (European Union);
A*MIDEX, ANR, IPhU and Labex P2IO, and R\'{e}gion Auvergne-Rh\^{o}ne-Alpes (France);
%RFBR, RSF and Yandex LLC (Russia);
Alexander-von-Humboldt Foundation (Germany);
ICSC (Italy); 
%GVA, XuntaGal, GENCAT, Inditex, InTalent and Prog.~Atracci\'on Talento, CM (Spain);
Severo Ochoa and Mar\'ia de Maeztu Units of Excellence, GVA, XuntaGal, GENCAT, InTalent-Inditex and Prog.~Atracci\'on Talento CM (Spain);
SRC (Sweden);
the Leverhulme Trust, the Royal Society, UKRI and IOP BBGSF (United Kingdom).

\addcontentsline{toc}{section}{References}

\ifx\mcitethebibliography\mciteundefinedmacro
\PackageError{LHCb.bst}{mciteplus.sty has not been loaded}
{This bibstyle requires the use of the mciteplus package.}\fi
\providecommand{\href}[2]{#2}

\newpage

% LHCb collaboration author list
% Data extracted on September 29th, 2025 at 6:53pm for paper reference LHCb-PAPER-2025-018
\centerline
{\large\bf LHCb collaboration}
\begin
{flushleft}
\small
R.~Aaij$^{38}$\lhcborcid{0000-0003-0533-1952},
A.S.W.~Abdelmotteleb$^{57}$\lhcborcid{0000-0001-7905-0542},
C.~Abellan~Beteta$^{51}$\lhcborcid{0009-0009-0869-6798},
F.~Abudin{\'e}n$^{57}$\lhcborcid{0000-0002-6737-3528},
T.~Ackernley$^{61}$\lhcborcid{0000-0002-5951-3498},
A. A. ~Adefisoye$^{69}$\lhcborcid{0000-0003-2448-1550},
B.~Adeva$^{47}$\lhcborcid{0000-0001-9756-3712},
M.~Adinolfi$^{55}$\lhcborcid{0000-0002-1326-1264},
P.~Adlarson$^{84}$\lhcborcid{0000-0001-6280-3851},
C.~Agapopoulou$^{14}$\lhcborcid{0000-0002-2368-0147}
% ...

{\footnotesize \it

% ...
$^{14}$Universit{\'e} Paris-Saclay, CNRS/IN2P3, IJCLab, Orsay, France\\
% ...
$^{38}$Nikhef National Institute for Subatomic Physics, Amsterdam, Netherlands\\
% ...
$^{47}$Instituto Galego de F{\'\i}sica de Altas Enerx{\'\i}as (IGFAE), Universidade de Santiago de Compostela, Santiago de Compostela, Spain\\
% ...
$^{51}$Physik-Institut, Universit{\"a}t Z{\"u}rich, Z{\"u}rich, Switzerland\\
% ...
$^{55}$H.H. Wills Physics Laboratory, University of Bristol, Bristol, United Kingdom\\
% ...
$^{57}$Department of Physics, University of Warwick, Coventry, United Kingdom\\
% ...
$^{60}$School of Physics and Astronomy, University of Glasgow, Glasgow, United Kingdom\\
$^{61}$Oliver Lodge Laboratory, University of Liverpool, Liverpool, United Kingdom\\
% ...
$^{69}$Syracuse University, Syracuse, NY, United States\\
% ...
$^{84}$Department of Physics and Astronomy, Uppsala University, Uppsala, Sweden, associated to $^{60}$\\
% ...
}
\end{flushleft}

\end{document}